\documentclass[aps,prl,twocolumn,superscriptaddress,showpacs]{revtex4}
\usepackage{amsfonts,amssymb,amscd,amsmath}
\usepackage{graphicx}
\usepackage{epsfig}
\usepackage{latexsym}
\usepackage{amssymb}
\usepackage{subfigure}
\usepackage{color}

\begin{document}

\def\abs#1{ \left| #1 \right| }
\def\lg#1{ | #1 \rangle }
\def\rg#1{ \langle #1 | }
\def\lrg#1#2#3{ \langle #1 | #2 | #3 \rangle }
\def\lr#1#2{ \langle #1 | #2 \rangle }
\def\me#1{ \langle #1 \rangle }

\newcommand{\bra}[1]{\left\langle #1 \right\vert}
\newcommand{\ket}[1]{\left\vert #1 \right\rangle}
\newcommand{\bx}{\begin{matrix}}
\newcommand{\ex}{\end{matrix}}
\newcommand{\be}{\begin{eqnarray}}
\newcommand{\ee}{\end{eqnarray}}
\newcommand{\nn}{\nonumber \\}
\newcommand{\no}{\nonumber}
\newcommand{\de}{\delta}
\newcommand{\lt}{\left\{}
\newcommand{\rt}{\right\}}
\newcommand{\lx}{\left(}
\newcommand{\rx}{\right)}
\newcommand{\lz}{\left[}
\newcommand{\rz}{\right]}
\newcommand{\inx}{\int d^4 x}
\newcommand{\pu}{\partial_{\mu}}
\newcommand{\pv}{\partial_{\nu}}
\newcommand{\au}{A_{\mu}}
\newcommand{\av}{A_{\nu}}
\newcommand{\p}{\partial}
\newcommand{\ts}{\times}
\newcommand{\ld}{\lambda}
\newcommand{\al}{\alpha}
\newcommand{\bt}{\beta}
\newcommand{\ga}{\gamma}
\newcommand{\si}{\sigma}
\newcommand{\ep}{\varepsilon}
\newcommand{\vp}{\varphi}
\newcommand{\zt}{\mathrm}
\newcommand{\bb}{\mathbf}
\newcommand{\dg}{\dagger}
\newcommand{\og}{\omega}
\newcommand{\Ld}{\Lambda}
\newcommand{\m}{\mathcal}
\newcommand{\dm}{{(k)}}

\title{Quantum noise limit for force sensitivity of linear detectors}
\author{Yang Gao}
\affiliation{Department of Physics,
Xinyang  Normal University, Xinyang, Henan 464000, China}
\author{Hopui Ho}
\author{Haixi Zhang}
\email{hxzhang@ee.cuhk.edu.hk} \affiliation{Department of Electronic
Engineering, The Chinese University of Hong Kong, \\ Shatin, NT,
Hong Kong SAR 852, China }

\begin{abstract}
We prove that the force sensitivity of the conventional
optomechanical detector associated with the optical quadrature
measurement of the output beam is lower bounded by the so-called
ultimate quantum limit (UQL), i.e., the absolute value of the
imaginary part of the inverse mechanical susceptibility. Through the
linear response theory, we find that the force sensitivity of any
linear detector is lower bounded by a generalized UQL, which might
beat the usual UQL by properly tailoring the detector-oscillator
interaction. We believe that our results open a new direction for
improving the performance of high-sensitivity detection schemes.
\end{abstract}

\maketitle


{\it Introduction.---}Quantum noise is known to impose fundamental
limits on high-sensitivity measurements \cite{qm,cave}. For a force
measurement with an optomechanical detector \cite{cave,clerk}, the
force is estimated from its effect on the position of a harmonic
oscillator. The displacement of the oscillator is then read out by a
probing laser beam. The force sensitivity of the measurement is
limited by two types of quantum noise: the shot noise of the laser
beam at the detection port and the radiation pressure backaction
noise introduced by the oscillator \cite{clerk}. An optimal tradeoff
between these noises induces a lower bound for classical detection
sensitivity, which is the so-called standard quantum limit (SQL)
\cite{qm,cave,clerk}.

However, the SQL itself is not a fundamental quantum limit. Various
schemes to overcome the SQL in force measurements have been
proposed, such as frequency dependent squeezing (FDS) of the input
beam \cite{fds,uql}, cavity detuning (CD) \cite{cd}, variational
measurement (VM) \cite{vm}, coherent quantum noise cancelation
(CQNC) \cite{cqnc}, etc. More importantly, an immediate question is
to find out the fundamental quantum limit for the force sensitivity.
In this letter, we aim to answer this question. For the mentioned
schemes that beat the SQL, we find that the corresponding force
sensitivities are lower bounded by the so-called UQL \cite{uql,cd}.
It is related to the dissipation mechanism of the oscillator, via
the absolute value of the imaginary part of the inverse mechanical
susceptibility. Through the linear response theory, we prove that
the force sensitivity of any linear detector is lower bounded by a
generalized UQL, which can be achieved by properly tailoring the
detector so as to overcome the usual UQL. This lower bound also
holds for the cases with coherent quantum control and/or quantum
feedback \cite{cqc,qfb}. The purpose of this letter is to provide a
criterion for the sensitivity limit, just as the Heisenberg limit in
quantum optical phase estimation \cite{hl}, and stimulate some
promising approaches for improving the performance of
high-sensitivity detection schemes.



\begin{figure}[htbp]
\centering \subfigure[]{
\includegraphics[angle=0,width=0.6\columnwidth]{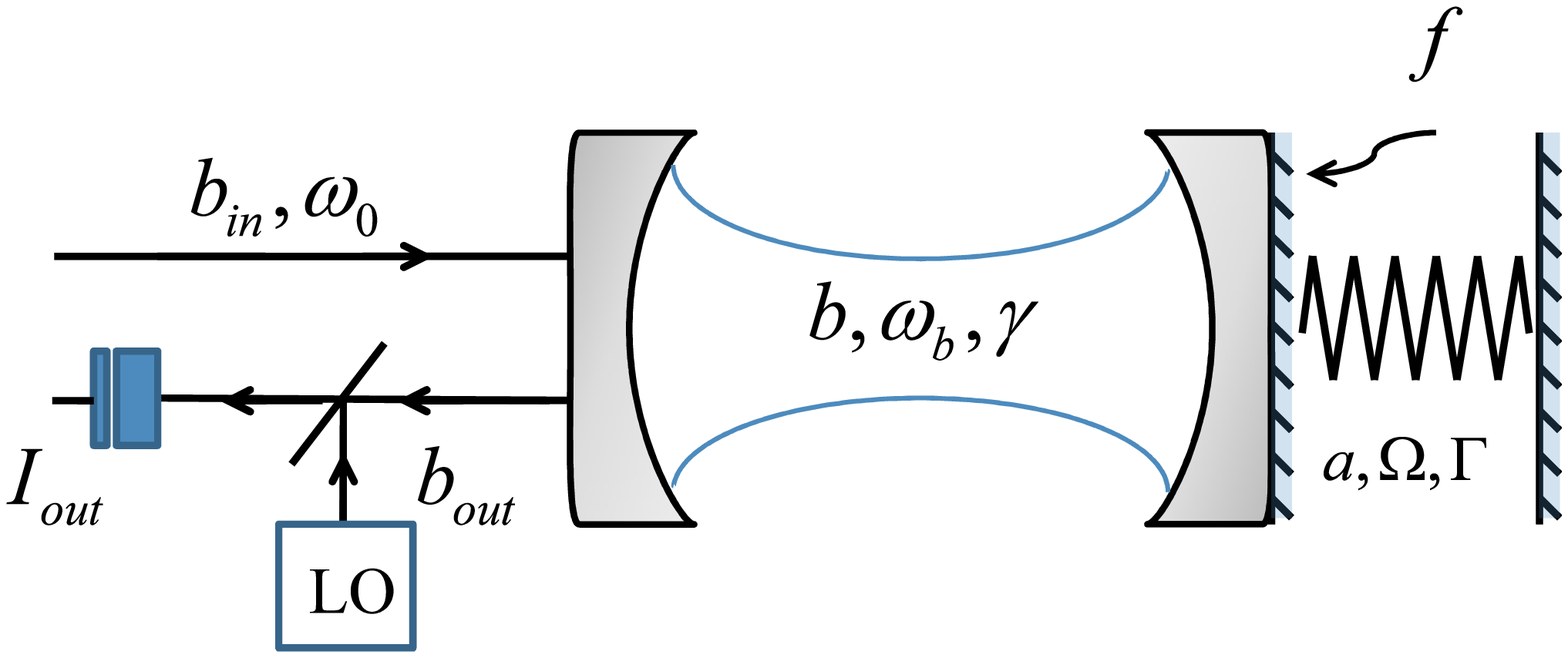}}\hspace{0.01in}
\subfigure[]{
\includegraphics[angle=0,width=0.6\columnwidth]{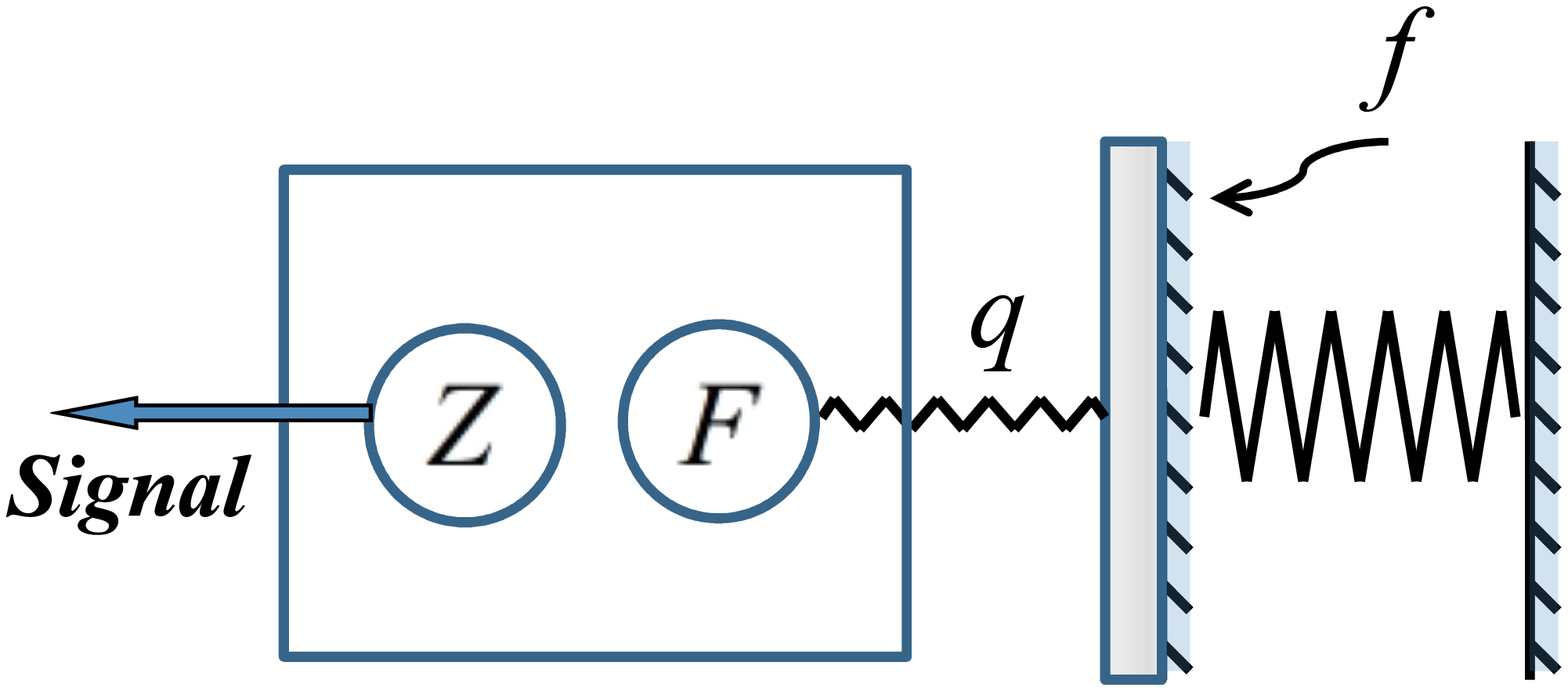}}\hspace{0.01in}
\caption{Schematics of the conventional optomechanical detector (a)
and the generic force detector (b). (a) The output current $I_{out}$
of the photodiode is modulated by the local oscillator phase. (b)
The detector's input and output operators $F$, $Z$ are coupled to
the oscillator and the readout apparatus, respectively.}
\end{figure}

{\it Optomechanical detector.---}The optomechanical detector
consists of a high quality Fabry-Perot cavity, with a fixed
transmissive mirror in front of the cavity, and a moveable,
perfectly reflecting mirror at the back (see Fig. 1a). The cavity is
in thermal equilibrium with the radiation, and is fed with a driving
laser. We aim to estimate a classical force acting on the moveable
mirror of cavity, for example, the passing of a gravitational wave
\cite{gwave}. In the rotating frame at the driving frequency $\og_0$
of the input laser, the system is described by the Hamiltonian
($\hbar=1$) \be H=H_m+H_o-g_{om}x (b^\dg b-\me{b^\dg b})+H_{dr}, \ee
where $H_m=\Omega a^\dg a-x f(t)$ is the mechanical oscillator under
the classical force $f(t)$, and $H_o= \Delta b^\dg b$ is the cavity
with the resonance frequency $\og_b$ at the equilibrium position
$x=0$ in the presence of the mean radiation pressure and the cavity
detuning $\Delta=\og_b-\og_0$. The third term captures
optomechanical interaction with the coupling strength $g_{om}$. The
last term $H_{dr}=i\sqrt{\ga}(\bt_{in}b^\dg-h.c.)$ is the laser
driving Hamiltonian. Taking into account the thermal noises, the
equations of motion are given by the quantum Langevin equations
\cite{qop}, \be \dot a &=&
i[H,a]-\frac{\Gamma}{2}a+\sqrt{\Gamma}a_{in}, \nn \dot b &=&
i[H,b]-\frac{\ga}{2}b+\sqrt{\ga}b_{in}, \label{lange} \ee where
$\Gamma (\ga)$ and $a_{in} (b_{in})$ are the decay rate and thermal
noise operator for the oscillator (cavity), respectively. The noise
correlators are $\me {a_{in}(t)a_{in}^\dg(t')}=(n_{th}+1) \de
(t-t')$ and $\me {b_{in}(t)b_{in}^\dg(t')}= \de (t-t')$, where
$n_{th}$ is the thermal occupancy of the mechanical reservoir.

Under the condition of strong laser driving, we can linearize the
system dynamics by splitting $a \to \me a + a$ and $b \to \me b +
b$, where $\me a=0$ and $\me b=\bt$ are the mean field values. The
linearized equation of motion is obtained by neglecting the
nonlinear terms in Eq. ({\ref{lange}}), \be \dot {\bb x} = {\bb A
\bb x}+\bb w, \label{eom} \ee where the variables $\bb
x=(x,p,b_1,b_2)^T$ with $a=(x+i p)/\sqrt{2}$ and $b=(b_1+i
b_2)/\sqrt{2}$, the input $\bb
w=(\sqrt{\Gamma}x_{in},f+\sqrt{\Gamma}p_{in},
\sqrt{\gamma}b_1^{in},\sqrt{\gamma}b_2^{in})^T$, and the matrix \be
\bb A= \lx \bx -{\Gamma \over 2} & \Omega & 0 & 0 \\ -\Omega &
-{\Gamma \over 2} & g & 0 \\ 0 & 0 & -{\gamma \over 2} & \Delta \\ g
& 0 & -\Delta & -{\gamma \over 2} \ex \rx, \ee where $g=\sqrt 2 \bt
g_{om} $ is the effective optomechanical coupling strength.

The classical force is estimated from the output current $I_{out}$
of a photodiode that is linearly proportional to a certain optical
quadrature of the output field, $ I_{out} \propto y \equiv
b_1^{out}\sin \phi +b_2^{out}\cos\phi$, where $\phi$ is the
adjustable readout quadrature angle via the local oscillator phase.
The output vector $\mathbf{b}_{out}=(b_1,b_2)^T_{out}$ in the
frequency domain, neglecting the intrinsic mechanical noise of the
oscillator, is determined by the input-output relation: \be
\mathbf{b}_{out} = {\bf M } \mathbf{b}_{in} + \mathbf{v} f,
\label{inout} \ee where $\bf M$ is the $2\times 2$ transfer matrix,
and ${\bf v}=(v_1,v_2)^T$. Putting Eq. (\ref{inout}) into the
expression of $y$, we get $ y = \mathbf{d}^T_\phi
\mathbf{b}_{out}=\mathbf{d}^T_\phi {\bf M}{\bf b}_{in}+
\mathbf{d}^T_\phi {\bf v} f$ with ${\bf d}=(\sin\phi,\cos\phi)^T$.
The first term represents the quantum noise, and the second term is
the output response to the classical force. The normalized
quadrature gives an unbias estimator $\hat{f}$ of $f$, $ \hat{f} =
{y}/(\mathbf{d}^T_\phi {\bf v})= f+f_{add}$, where
$f_{add}=\mathbf{d}^T_\phi {\bf M}{\bf b}_{in}/(\mathbf{d}^T_\phi
{\bf v})$ is the added noise. The force sensitivity is quantified by
the power density of the added noise \be S_f
\de(\og-\og')=\me{f_{add}(\og)f^\dg_{add}(\og')
+f^\dg_{add}(\og')f_{add}(\og)}/2. \ee It gives $
S_f={\mathbf{d}^T_\phi {\bf M} {\bf S} {\bf M}^\dg
\mathbf{d}_\phi}/{|\mathbf{d}^T_\phi {\bf v}|^2}$, where the power
density matrix $\bf S$ for the quadratures $b_1^{in}$ and $b_2^{in}$
is defined by
$S_{XY}(\og)\de(\og-\og')=\me{X(\og)Y^\dg(\og')+Y^\dg(\og')X(\og)}/2$.

For a resonant cavity ($\Delta=0$), the added noise is \cite{supple}
\be f_{add} =\frac{\xi b_1^{in} +b_2^{in}}{g \sqrt{\ga}\chi_a
\chi_b^*}+ g \sqrt{\ga} \chi_b b_1^{in}, \label{add} \ee where
$\chi_a={\Omega}\lz{(\Gamma/2-i\og)^2+\Omega^2}\rz^{-1}$, $\chi_b =
\lx{\ga/2-i \og}\rx^{-1}$, and $\xi=\tan \phi$. Here the first term
is the shot noise at the detection port, and the second term is the
backaction noise from the oscillator due to radiation pressure. For
non-squeezed coherent input laser, ${\bf S}={\bf I}/2$. Assuming
$\phi=\xi=0$, Eq. (\ref{add}) gives rise to \be S_f=\frac{1}{2}\lx
{g^2\ga}|\chi_b|^2+\frac{1}{g^2\ga |\chi_a|^2 |\chi_b|^2} \rx \ge
\frac{1}{|\chi_a|}, \ee where the inequality $a+b \ge 2\sqrt{|ab|}$
has been used. The minimal sensitivity is achieved when the
backaction noise and shot noise are balanced, known as the SQL for
the detector sensitivity \cite{qm,cave,clerk}. However, the SQL can
be overcome by several schemes.


{\it Schemes to beat the SQL.---}A simple scheme to beat the SQL is
just varying the readout quadrature angle ($\phi \neq 0$) \cite{vm}.
It introduces an extra $\xi$-dependent shot noise to destructively
interfere with the original backaction noise and to give a better
sensitivity. From Eq. (\ref{add}), we have in terms of
$\bar{\chi}_a=1/{\chi}_a=\bar{\chi}_a^R+i\bar{\chi}_a^I$, \be S_f
&=& \xi\bar{\chi}_a^R+\frac{1}{2}\lx
{g^2\ga}|\chi_b|^2+\frac{1+\xi^2}{g^2\ga |\chi_a|^2 |\chi_b|^2} \rx
\nn &\ge& \xi\bar{\chi}_a^R + \frac{\sqrt{1+\xi^2}}{|\chi_a|} \ge
|\bar{\chi}_a^I|, \ee where the inequalities $a+b \ge 2\sqrt{|ab|}$
and $a x +\sqrt{(1+x^2)(a^2+b^2)} \ge |b|$ have been used. This
lower bound for force sensitivity is the known UQL in Refs.
\cite{cd,uql}, which has also been discussed in the limits of weak
coupling and high power gain in Ref. \cite{clerk}.

Another scheme to beat the SQL in Ref. \cite{fds} is to use a
frequency dependent squeezed input laser with the elements, $S_{11}=
u$, $S_{22}=v$, and $S_{12}=S_{21}=w$. The interference between the
backaction noise and the shot noise due to the correlation between
$b_1^{in}$ and $b_2^{in}$ ($w\neq 0$) could give rise to certain
negative terms in the $S_f$, and thus surpass the SQL. For a coupled
VM-FDS scheme, Eq. (\ref{add}) gives \be S_f &=& 2 (\xi
u+w)\bar{\chi}_a^R+{g^2\ga |\chi_b|^2} u + \frac{\xi^2 u+2\xi w+v
}{g^2\ga |\chi_a|^2|\chi_b|^2} \nn &\ge& 2 (\xi u+w)\bar{\chi}_a^R
+\frac{2}{|\chi_a|} \sqrt{u(\xi^2 u+2\xi w+v)} \nn &\ge& h
\bar{\chi}_a^R +\frac{\sqrt{1+h^2}}{|\chi_a|} \ge |\bar{\chi}_a^I|,
\quad h=2 (\xi u+w), \label{sim} \ee where the inequalities $a+b\ge
2\sqrt{|ab|}$, $ uv-w^2 \ge {1/4}$, and $a x
+\sqrt{(1+x^2)(a^2+b^2)} \ge |b|$ have been used. In Ref. \cite{cd},
a nonzero cavity detuning ($\Delta \neq 0$) is invoked, which
simultaneously modifies the backaction noise and the shot noise, to
beat the SQL. As shown in Fig. 2a, the UQL sets a lower bound to all
the above schemes. Remarkably, the force sensitivity for a combined
VM-FDS-CD scheme is also lower bounded by the UQL \cite{supple}.

Finally, we show that the CQNC scheme \cite{cqnc} that could beat
the SQL still satisfies the UQL. In the CQNC scheme, an additional
ancillary cavity of mode $c(t)=(c_1+ic_2)/\sqrt 2$ fed with the
vacuum is introduced, and the following Hamiltonian is assumed: \be
H_c=-\Omega c^\dg c-gb_1 c_1. \label{anc} \ee The ancilla works
effectively as a negative-mass oscillator, and its coupling with the
main cavity can be realized via a beam-splitter and a nondegenerate
optical parametric amplifier. Considering the intrinsic mechanical
and cavity noises, the dynamics of the system is also governed by
Eq.(\ref{eom}), where the variables $\bb x=(x,p,b_1,b_2,c_1,c_2)^T$,
the input $\bb w=(\sqrt{\Gamma}x_{in},f+\sqrt{\Gamma}p_{in},
\sqrt{\gamma}b_1^{in},\sqrt{\gamma}b_2^{in},
\sqrt{\Gamma}c_1^{in},\sqrt{\Gamma}c_2^{in})^T$, and the matrix $\bb
A$ is given in \cite{supple}. Based on the output photoncurrent at
readout angle $\phi$, the added noise takes \be f_{add}= {\xi
b_1^{in}+b_2^{in} \over g\sqrt{\ga}\chi_a\chi_b^*}+\m F_c, \ee where
$\m F_c={\sqrt{\Gamma}}\lz c_1^{in}(\Gamma/2-i\og)/\Omega -c_2^{in}
\rz$ is the intrinsic noise from the ancillary cavity. It can be
seen that the backaction noise from the main cavity was canceled
out. For sufficiently large pump $g \gg 1$, the shot noise from the
main cavity can be made insignificant with respect to $\m F_c$. The
detector sensitivity is essentially given by the power density of
$\mathcal F_c$, \be S_f \approx
\frac{\Gamma}{2\Omega^2}\bigg(\og^2+\Omega^2+{\Gamma^2 \over
4}\bigg) \ge {\og \Gamma \over \Omega} = |\bar \chi_a^I|,\ee namely,
the UQL.

{\it Optimal force sensitivity for linear response detector.---} For
the conventional optomechanical detector, the above results suggest
that the UQL might be true for any detection scheme, such as the
cases in Ref. \cite{more,pt}. To determine this conjecture, we
consider some physical system as a generic linear response detector
(see Fig. 1b). The detector is described by some unspecified
Hamiltonian $H_d$, and has both an input operator, represented by an
operator $F$, and an output operator, represented by an operator $\m
Z$. The input operator $F$ is coupled with the mechanical oscillator
via the interaction Hamiltonian, $H_{int}=-g F q$, where the
oscillator operator $q$ is not necessarily the position operator
$x$, as long as it carries the input signal. The output operator $\m
Z$ (e.g., the output optical quadrature $y$) is related to the
readout quantity at the output of the detector (e.g., the output
current $I_{out}$ of the photodiode), from which the classical force
is estimated.

The total Hamiltonian is given by $H=H_m+H_d+H_{int}$. Treating
$H_{int}$ as the perturbation, an arbitrary operator $O$ in the
Heisenberg picture is obtained by \be O(t)= \mathcal U^\dg O_0(t)
\mathcal U, \quad \mathcal U(t)=\mathcal T e^{-i\int_{-\infty}^t
H^0_{int}(t) dt}, \label{his} \ee where $O_0(t)$ denotes the
operator in the interaction picture, and the symbol $\mathcal T$
means the time-ordered product. For the linear operators $q$, $F$,
and $Z$ (with the c-number commutators), Eq. (\ref{his}) gives
\cite{supple} \be q(\og) &=& q_0(\og)+ \chi_{qx}(\og)f(\og)+ g
\chi_{qq}(\og) F(\og), \nn F(\og) &=& F_0(\og)+ g \chi_{F\!\!\!\
F}(\og) q(\og), \nn Z(\og) &=& Z_0(\og)+ g q(\og), \label{lre} \ee
where $Z$ is the rescaled output operator via $\m Z(\og)=\chi_{\m
Z\!\!\!\ F}(\og) Z(\og) $. The susceptibility $\chi_{XY}$ is defined
via the c-number commutator, $ \chi_{X \!\!\!\ Y}(t)=i
\theta(t)[X(t),Y(0)]$.

Solving the first two equations and substituting into the third one
of Eq. (\ref{lre}), we have the output operator in terms of the
unperturbed operators, \be Z(\og) = Z_0+g{q_0+\chi_{qx}f+ g\chi_{qq}
F_0 \over 1 - g^2\chi_{qq}\chi_{F\!\!\!\ F}}. \ee The normalization
of $Z$ gives the estimator $\hat f$ of $f$, \be \hat f = f + \bar
\chi_{qx} (q_0+ G_F F_0+G_Z Z_0)=f+f_{add}, \label{add2} \ee where
$G_F= g \chi_{qq}$ and $G_Z=(1 -g^2 \chi_{qq} \chi_{F\!\!\!\ F})/g$.
The $F_0$-term represents the backaction noise from the oscillator,
while the $Z_0$-term is the shot noise at the output. Neglecting the
intrinsic mechanical noise $\propto q_0$, the scaled power density
$S'_f = S_f|\chi_{qx}|^2$ takes $ S'_f = |G_F|^2 S_{F\!\!\!\
F}+|G_Z|^2 S_{Z\!\!\!\ Z} + (G_FG_Z^* S^*_{Z\!\!\!\ F}+G_F^*G_Z
S_{Z\!\!\!\ F})$, where the relation $S_{F\!\!\!\ Z}=S^*_{Z\!\!\!\
F}$ has been used. The optimization of $S'_f$ over the coupling
strength $g$ gives \be S'_f &\ge& 2 (A \chi^R_{qq} - B \chi^I_{qq})
+ |\chi_{qq}|  \big[ S_{Z\!\!\!\ Z}(S_{F\!\!\!\ F} + |\chi_{F\!\!\!\
F}|^2 S_{Z\!\!\!\ Z} \nn && - 2 \chi^R_{F\!\!\!\ F} S^R_{Z\!\!\!\ F}
+ 2 \chi^I_{F\!\!\!\ F} S^I_{Z\!\!\!\ F})\big]^{1/2}, \label{ieq1}
\ee where $A=-\chi^R_{F\!\!\!\ F} S_{Z\!\!\!\ Z} + S^R_{Z\!\!\!\ F}$
and $B=\chi^I_{F\!\!\!\ F} S_{Z\!\!\!\ Z}+S^I_{Z\!\!\!\ F}$.

To proceed further, we must at least require $ [\m Z(t),\m
Z(t')]=[\m Z_0(t),\m Z_0(t')]=0 $ at all times, in order for $\m
Z(t)$ and $\m Z_0(t)$ to represent experimental data string. It
immediately implies that $\chi_{\m Z\!\!\!\ \m Z}(\og)=0$. Also, the
causality principle imposes that the output $\m Z_0(t)$ should not
depend on the input $F_0(t')$ for $t<t'$, and therefore
$\chi_{F\!\!\!\ \m Z}(\og)=0$. Furthermore, $F$ and $Z$ should
satisfy the uncertainty relation \cite{supple}, \be S_{F\!\!\!\
F}S_{Z\!\!\!\ Z} \ge |S_{Z\!\!\!\ F}|^2+|B|+1/4. \label{ieq2} \ee

Putting Eq. (\ref{ieq2}) into (\ref{ieq1}), we obtain \be S'_f &\ge&
2 (A \chi^R_{qq} - B\chi^I_{qq}) + 2|\chi_{qq}|
\sqrt{A^2+(|B|+1/2)^2} \nn &\ge& -2B\chi^I_{qq}
+2(|B|+1/2)|\chi^I_{qq}| \ge |\chi^I_{qq}|, \ee where the
inequalities $a_1 x_1 +\sqrt{(a_1^2+a_2^2)(x_1^2+x_2^2)} \ge |a_2
x_2|$ and $|a| \ge a$ have been applied. The resulting force
sensitivity is \be S_f \ge {|\chi^I_{qq}|}/{|\chi_{qx}|^2},
\label{main} \ee which is the main result of this letter. It can be
viewed as a generalized UQL. For the optomechanical detector, we
have $q=x$, $F=b_1$, $\m Z=y$, $\chi_{qx}=\chi_{qq}=\chi_a$, and
thus the usual UQL for the detector sensitivity. As an example, we
could conclude that the sensitivity with a PT-symmetric cavity near
the PT-phase transition \cite{pt} can not be enhanced below the
usual UQL, because therein the optomechanical interaction is of the
form $-g x F$.

\begin{figure}[t!]
\centering \subfigure[]{
\includegraphics[width=0.6\columnwidth]{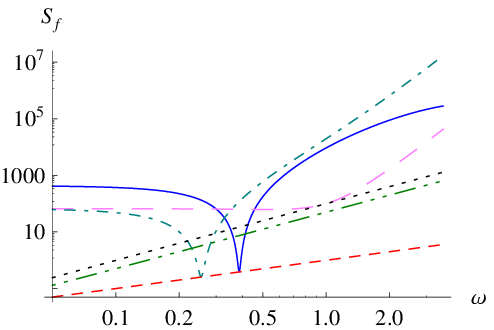}}\hspace{0.01in}
\subfigure[]{
\includegraphics[width=0.6\columnwidth]{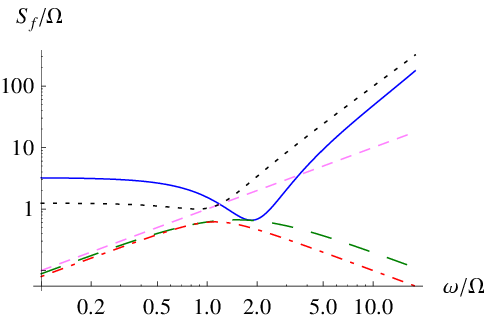}}\hspace{0.01in}
\caption{Plot of the force sensitivity versus the detection
frequency. (a) Solid line: CD scheme ($\Delta=-7$); dot-dashed line:
VM scheme ($\xi=20$); long-dashed line: the standard detection
scheme; double-dot-dashed line: CQNC scheme; dotted line: the SQL;
dashed line: the UQL. Here the system parameters are
$\mathbf{S}=\mathbf{I}/2$, $\gamma =3$, $\Gamma = \Omega=0.01$, and
$g=-10$. (b) Solid line: the toy optomechanical detector;
long-dashed line: the generalized UQL ($q=x+p$); dashed line: the
usual UQL ($q=x$); dotted line: the SQL; dot-dashed line: the
optimal UQL. Here the system parameters are
$\mathbf{S}=\mathbf{I}/2$, $\gamma = 100 \Omega$, $\Gamma = \Omega$,
$g = 5\Omega$, and $\xi = 0$.}
\end{figure}

On the other hand, if the detector is coupled with the oscillator
via a certain $q \neq x$, the lower bound in Eq. (\ref{main}) might
be achieved by tuning the detector structure and even beat the usual
UQL. It can be understood that the backaction noise and the shot
noise are modified with more freedoms to fulfill a destructive
interference in the $S_f$. As an example, we devise a toy
optomechanical detector. The cavity-oscillator interaction is
supposed to be $H_{int}=-g (x+p)(b_1+b_2)$. The
matrix $\bb A$ becomes \be \bb A= \lx \bx -{\Gamma \over 2} & \Omega & -g & -g \\
-\Omega & -{\Gamma \over 2} & g & g \\ -g & -g & -{\gamma \over 2} &
0 \\ g & g & 0 & -{\gamma \over 2} \ex \rx, \label{super} \ee which
satisfies the stability condition since the eigenvalues of $\bb A$
are ($-{\gamma \over 2}$, $-{\gamma \over 2}$, $-{\Gamma \over 2}+
i\Omega$, $-{\Gamma \over 2} - i\Omega$). This type of coupling
might be realized in one-dimensional superconducting stripline
resonators \cite{bu}. Based on the measurement of $y$, the power
density of the added noise can be calculated. The numerical result
is plotted in Fig. 2b. It shows that the generalized UQL \be S_f \ge
2\lz\bigg(1+\frac{\Gamma}{2\Omega}\bigg)^{2}+\frac{\og^2}{\Omega^2}\rz^{-1}
{\omega \Gamma \over \Omega} \label{guql} \ee is achievable at a
certain frequency and beats the usual UQL. Eq. (\ref{guql}) is
obtained from Eq. (\ref{main}) with the susceptibilities
$\chi_{qq}=2\chi_a$ and $\chi_{qx}=\lz 1+(\Gamma/2-i \og)/\Omega\rz
\chi_a$. Moreover, by varying over all possible linear coupling
operator $q=x + \eta p$ in Eq. (\ref{main}), we get the optimal UQL
\be S_f &\ge& {\Gamma \over 2 \og \Omega}
\bigg[\frac{\Gamma^2}{4}+\og^2+ \Omega^2 \nn && -\sqrt{
\bigg(\frac{\Gamma^2}{4}+\omega^2-\Omega^2\bigg)^2+\Gamma^2\Omega^2}\bigg],
\label{optim} \ee which approaches $\Gamma\Omega/\og$ for $\og \gg
\Omega$, and is different from the usual UQL scaling as $\og
\Gamma/\Omega$.

Obviously, the above result has incorporated the effect of coherent
quantum control \cite{cqc}, such as the CNQC scheme. As for the
direct quantum feedback control \cite{qfb}, the output signal is fed
back to the system for changing the dynamical evolution. It
introduces an additional term to the equation of motion for a
generic operator $O$, $\dot O_{fb}(t) = i \ld \m Z(t) [\m P,O]$,
where $\ld$ is the feedback gain, $\m P(t)$ is the control operator.
Within this formalism, the generalized UQL for force sensitivity
still holds \cite{supple}.

{\it Conclusion.---}We have shown that the force sensitivity of the
standard optomechanical detector associated with the optical
quadrature measurement of the output field is lower bounded by the
usual UQL. By the linear response theory, we have also found that
the force sensitivity of any linear detector is lower bounded by the
generalized UQL, which can beat the usual UQL by appropriately
tailoring the detector. A toy optomechanical detector is devised to
beat the usual UQL. We believe that this study provides a criterion
for the sensitivity limit, just as the Heisenberg limit in quantum
optical phase estimation, and even gives a promising approach for
improving the performance of high-sensitivity detection schemes.

YG acknowledges the support from NSFC Grant No. 11304265. HH and HZ
acknowledge financial support from the CRF scheme (CUHK1/CRF/12G),
the ITF scheme (GHP/014/13SZ), and the AoE scheme (AoE/P-02/12) from
the Research Grants Council (RGC) of Hong Kong Special
Administrative Region, China.


\newpage
{\bf Supplemental Materials}

\subsection{Optomechanical detector}

The equation of motion of the optomechanical system is given by Eq.
(\ref{eom}). The explicit form takes \be \dot{x} &=&
-\frac{\Gamma}{2}x + \Omega p+\sqrt{\Gamma}x_{in}, \nn \dot p &=&
-\Omega x-\frac{\Gamma}{2}p+g_{om}(b^\dg b-\me {b^\dg
b})+f+\sqrt{\Gamma}p_{in}, \nn \dot b &=& -i\Delta b -\frac{\ga}{2}
b+i g_{om} x b+\sqrt{\ga}(\bt_{in}+b_{in}). \label{lin} \ee Then we
find the steady mean state to be $\me x =\me p=0$ and $\me
b=\bt=\sqrt{\ga}\bt_{in}/({\ga/2}+i\Delta)$. For convenience, we
have chosen $\bt$ as real by adjusting the driving field $\bt_{in}$.
The linearization of Eq. (\ref{lin}) around this steady state gives
Eq. (\ref{eom}). The stability of this linearized system is
guaranteed by the requirement that the real part of all the
eigenvalues of $\bb A$ must be nonpositive. For the stationary
state, the Fourier transform of Eq. (\ref{eom}) \be -i \og {\bf
x}={\bf A} {\bf x}+ {\bf w} \ee yields the solution ${\bf x}=-({\bf
A}+i \og {\bf I})^{-1} {\bf w}$. The output field is obtained by the
input-output relation $b_{out}=\sqrt{\kappa}b-b_{in}$. Neglecting
the intrinsic mechanical noises due to $x_{in}$ and $p_{in}$, we get
Eq. (\ref{inout}), where ${\bf M}={\bf M}_{shot}+{\bf M}_{back}$,
\be {\bf
M}_{shot}&=&-{\bf I}+ \ga \lx \bx \chi_r & \chi_\Delta \\
-\chi_\Delta & \chi_r \ex \rx, \nn {\bf M}_{back} &=& {g^2 \ga
\chi_a \over 1-g^2\chi_a\chi_\Delta}
\lx \bx \chi_r \chi_\Delta & \chi_\Delta^2 \\
\chi_r^2 & \chi_r \chi_\Delta \ex \rx, \nn {\bf v}&=& {g \sqrt {\ga}
\chi_a \over 1-g^2\chi_a\chi_\Delta} \lx \bx \chi_\Delta \\ \chi_r
\ex \rx. \ee Here we have separated the output field into the shot
noise term, i.e., the output field without the interaction ($g=0$),
and the backaction noise term depending on $g$. The quantities
$\chi_{r (\Delta)}$ are defined by \be
\chi_r&=&\frac{r}{r^2+\Delta^2}, \nn
\chi_\Delta&=&\frac{\Delta}{r^2+\Delta^2} \ee with $r=\ga/2-i\og$.
The added noise $f_{add}$ can then be written as \be
f_{add}=\mathbf{d}^T_\phi {\bf M}_{shot}{\bf
b}_{in}/(\mathbf{d}^T_\phi {\bf v})+\mathbf{d}^T_\phi {\bf
M}_{back}{\bf b}_{in}/(\mathbf{d}^T_\phi {\bf v}). \label{noise} \ee
For a resonant cavity ($\Delta=0$), $\chi_r=\chi_b$ and
$\chi_\Delta=0$, we get \be {\bf M}_{shot} &=& e^{i\theta} {\bf I},
\quad e^{i\theta}=\chi_b/\chi_b^*, \nn
{\bf M}_{back} &=& g^2 \ga \chi_a\chi_b^2\lx \bx 0 & 0 \\
1 & 0 \ex \rx, \nn {\bf v}&=&g \sqrt {\ga} \chi_a\chi_b \lx \bx 0 \\
1 \ex \rx. \ee Substituting them into Eq. (\ref{noise}) gives the
added noise in Eq. (\ref{add}).

To prove the UQL for the combined VM-FDS-CD scheme, we note that
$f_{add}$ obtained from Eq. (\ref{noise}) takes the form \be
f_{add}={\bar \chi_a D +E \over C} b_1^{in} + {\bar \chi_a X+ Y
\over C} b_2^{in}, \ee where \be D&=&\xi (|r|^2-\Delta^2)-\ga
\Delta, \nn E&=&g^2(\ga+\xi \Delta),\nn X&=&|r|^2-\Delta^2+\xi
\ga\Delta,\nn Y&=&g^2 \Delta, \nn C&=&g\sqrt{\ga}(r+\xi \Delta). \ee
Using the input density matrix $\bf S$ (with elements $S_{11}=u$,
$S_{22}=v$, and $S_{12}=S_{21}=w$), the sensitivity is given by \be
S_f = 2\bar \chi_a^R H + K + |\bar \chi_a|^2 L \nn \ge 2\bar
\chi_a^R H +2 |\bar \chi_a| \sqrt{KL}, \ee where \be H &=&
(DEu+EXw+DYw+XYv)/|C|^2, \nn K&=&(E^2 u+2 EYw+Y^2v)/|C|^2, \nn
L&=&(D^2u+2DXw+X^2v)/|C|^2. \ee It is simple to check the identities
\be KL-H^2&=&(uv-w^2)(EX-DY)^2/|C|^2, \nn (EX-DY)^2&=&|C|^2.\ee
Noting the inequality $uv-w^2 \ge 1/4$, we finally have \be S_f\ge
2\bar \chi_a^R H +2 |\bar \chi_a|\sqrt{H^2+1/4} \ge |\bar \chi_a^I|,
\ee which is the UQL for the optomechanical detector. For a pure
squeezed input driving, we have \be u&=&\cosh 2s-\sinh 2s \cos 2\vp,
\nn v&=&\cosh 2s+\sinh 2s \cos 2\vp, \nn w&=&-\sinh 2s \sin 2\vp,
\ee and $uv-w^2 = 1/4$, where $s$ is called the squeezing factor and
$\vp$ is the squeezing angle.

For the CQNC scheme, the ancillary Hamiltonian can be realized by a
beam splitter described by the interaction form $-g(bc^\dg+b^\dg
c)/\sqrt{2}$ plus a non degenerate optical parameter amplifier
described by the interaction term $-g(e^{-2 i\og_0 t}b^\dg
c^\dg+e^{2 i\og_0 t} b c)/\sqrt{2}$. In the rotating frame at
frequency $\og_0$ ($b \to e^{-i\og_0 t} b$ and $c \to e^{-i\og_0 t}c
$), the relevant Hamiltonian becomes $H_c=(\og_c-\og_0)c^\dg c-g
b_1c_1$. Choosing the detuning $\og_c-\og_0=-\Omega$, we get the
Hamiltonian in Eq. (\ref{anc}). So the ancilla works as a negative
mass oscillator in order to cancel the backaction noise from the
main cavity. The resulting matrix for the resonant case ($\Delta=0$)
is
\be \bb A= \lx \bx -{\Gamma \over 2} & \Omega & 0 & 0 & 0 & 0 \\
-\Omega
& -{\Gamma \over 2} & g & 0 & 0 & 0 \\ 0 & 0 & -{\gamma \over 2} & 0 & 0 & 0 \\
g & 0 & 0 & -{\gamma \over 2} & g & 0 \\ 0 & 0 & 0 & 0 & -{\Gamma \over 2} & -\Omega \\
0 & 0 & g & 0 & \Omega & -{\Gamma \over 2} \ex \rx, \ee where the
decay rate of the ancilla is assumed to be the same as the
mechanical oscillator. The final output field is given by \be {\bf
b}_{out} = e^{i\theta} {\bf b}_{in}+g \sqrt{\ga} \chi_a\chi_b \lx
\bx 0 \\1 \ex \rx (f+\m F_c), \ee where the backaction noise from
the main cavity vanishes, due to the coherent cancelation, at the
cost of an extra noise $\m F_c$ from the ancillary cavity.

\subsection{Linear system and spectral uncertainty relations}

The general linear system we described in the main text is
$H=H_m+H_d+H_{int}$ with $H_{int}=-g F q$. The operator in the
Heisenberg picture is given by Eq. (\ref{his}), where $H_{int}^0=-g
F_0 q_0$. Expanding the time-ordered exponential $\m U(t)$, we have
\be  \m U(t)&=& \mathbf I+{1\over i}\int_{-\infty}^t dt_1
H_{int}^0(t_1)+ {1\over i^2} \int_{-\infty}^t dt_1H_{int}^{0}(t_1)
\nn && \times \int_{-\infty}^{t_1}dt_2 H_{int}^{0}(t_2)+\cdots \nn
&=& \mathbf I+  {1\over i} \int_{-\infty}^t dt_1 H_{int}^{0}(t_1) \m
U(t_1), \ee and thus \be O(t)&=& \m U^\dg (t) O_0(t) \m U(t) \label{hiss} \\
&=& O_0(t) +{1\over i}\int_{-\infty}^t dt_1 \m U^\dg (t_1)
[O_0(t),H_{int}^0(t_1)]\m U(t_1). \no \ee For $O=q$, we note that
$[q_0(t),F_0(t')]=0$ since $q_0$ and $F_0$ are independent
variables, and $[q_0(t),q_0(t')]$ is a c-number for the linear
operator $q_0$. So Eq. (\ref{hiss}) gives \be q(t) &=&
q_0(t)+i\int_{-\infty}^tdt_1[q_0(t),x_0(t_1)]f(t_1)\nn &&
+ig\int_{-\infty}^t dt_1[q_0(t),q_0(t_1)]F(t_1), \label{pert} \ee
where the second term comes from the action of the external force,
and the relation $F(t_1)=\m U^\dg (t_1) F_0(t_1) \m U(t_1)$ has been
used. For a stationary system, we introduce the susceptibility \be
\chi_{XY}(t-t')=i\theta(t-t')[X(t),Y(t')].\ee The Fourier transform
of Eq. (\ref{pert}) immediately gives the first line of Eq.
(\ref{lre}). Similarly, the equations for $F$ and $Z$ can be
obtained.

Now we outline the proof of the spectral uncertainty relations for
arbitrary two linear Hermitian operators $O_1^0$ and $O_2^0$. Let us
consider an operator of the form \be P=
\sum_{j=1}^2\int_{-\infty}^\infty dt \zeta_j(t) O_j^0(t), \ee where
$\zeta_j$ are arbitrary complex functions. The positivity of the
Hermitian operator $\me {O^\dg O}$ implies that
\be\sum_{j,k}\int_{-\infty}^\infty dt \int_{-\infty}^\infty dt'
\zeta^*_j(t) \zeta_k(t') \me {O_j^{0}(t) O_k^0(t')} \ge 0 .
\label{s2} \ee We note the identity \be \me {O_j^0(t)
O_k^0(t')}=S_{jk}(t-t')+[O_j^0(t), O_k^0(t')]/2 \nn =
S_{jk}(t-t')-i[\chi_{jk}(t-t')-\chi_{kj}(t'-t)]/2, \ee where the
symmetrized correlator \be S_{jk}(t-t')= \me {O_j^0(t)
O_k^0(t')+O_k^0(t')O_j^0(t) }/2 \ee is related to the power density
$S_{jk}(\og)$ via the Fourier transform
$S_{jk}(t)=\int_{-\infty}^\infty S_{jk}(\og)e^{-i\og t}d\og$. In the
frequency domain, Eq. (\ref{s2}) becomes \be
\sum_{j,k}\int_{-\infty}^\infty d\og \zeta^*_j(\og)
M_{jk}(\og)\zeta_k(\og) \ge 0 \ee with the notation \be M_{jk}(\og)
= S_{jk}(\og)-i[\chi_{jk}(\og)-\chi_{kj}^*(\og)]/2. \ee It implies
that the $2\times 2$ the Hermitian matrix $M_{jk}$ is positive. This
is equivalent to the following three spectral uncertainty relations
$S_{11}(\og) \ge -\chi_{11}^I(\og)$, $S_{22}(\og) \ge
-\chi_{22}^I(\og)$, and
 \be [S_{11}(\og) +\chi_{11}^I(\og)][S_{22}(\og) +\chi_{22}^I(\og)] \nn \ge
|S_{21}-i[\chi_{21}(\og)-\chi_{12}^*(\og)]/2|^2. \ee Following the
similar arguments for the positivity of $\me {O O^\dg}$, we obtain
$S_{11}(\og) \ge \chi_{11}^I(\og)$, $S_{22}(\og) \ge
\chi_{22}^I(\og)$, and \be [S_{11}(\og) -
\chi_{11}^I(\og)][S_{22}(\og) - \chi_{22}^I(\og)] \nn \ge
|S_{21}+i[\chi_{21}(\og)-\chi_{12}^*(\og)]/2|^2. \ee

For the case of $O_1=F$, $O_2=Z$, since $\chi_\m {ZZ}=\chi_{F\m
Z}=0$ and $[\m Z(\og),F(\og')]=i\delta(\og+\og')[\chi_{\m
ZF}(\og)-\chi_{F\m Z}^*(\og)]$ imply $\chi_{ZZ}=\chi_{FZ}=0$ and
$\chi_{ZF}-\chi_{FZ}^*=1$.  The above spectral uncertainty relations
lead to $S_{FF} \ge |\chi_{FF}^I|$, $S_{ZZ} \ge 0$, and \be
(S_{FF}\pm \chi_{FF}^I) S_{ZZ} \ge |S_{ZF}\mp i/2|^2. \ee They can
be put into a more succinct form \be S_{FF}S_{ZZ} \ge
|S_{ZF}|^2+|B|+1/4 \ee with $B=\chi_{FF}^I S_{ZZ}+S^I_{ZF}$, which
is just Eq. (\ref{ieq2}) in the main text.

\subsection{Optimal UQL}

In order to optimize the generalized UQL, \be S_f\ge
|\chi_{qq}^I|/|\chi_{qx}|^2, \ee over all possible $q=x + \eta p$,
we need the expressions for $\chi_{qq}$ and $\chi_{qx}$. They can be
derived through $H_m \to H_m-f_q q-f_x x$ and the relation
$q=\chi_{qq}f_q+\chi_{qx}f_x$.
The equation of motion gives \be \lx \bx {\Gamma \over 2}-i\og & -\Omega \\
\Omega & {\Gamma \over 2}-i\og \ex \rx \lx \bx x \\ p \ex \rx = \lx \bx -\eta f_q \\
f_q+f_x \ex \rx, \ee and thus \be q=x+\eta p = (1+\eta^2) \chi_a
f_q+ \lx 1+\eta {\Gamma/2-i\og \over \Omega} \rx \chi_a f_x. \no \ee
So we read \be  \chi_{qq}&=& (1+\eta^2) \chi_a, \nn \chi_{qx} &=&\lx
1+\eta {\Gamma/2-i\og \over \Omega} \rx \chi_a. \ee Substituting
them into Eq. (\ref{main}), we get \be S_f \ge \frac{(1+\eta^2)\og
\Gamma/\Omega}{[1+\eta \Gamma/(2\Omega)]^2+(\eta \og /\Omega)^2}.
\ee The optimization over $\eta$ gives Eq. (\ref{optim}), i.e., the
optimal UQL.

\subsection{Direct quantum feedback control}

The direct quantum feedback control feeds the output signal back to
the original system for changing the dynamical evolution. It can be
represented by an additional term for the equation of motion of a
generic operator, \be \dot {\m O}_{fb}(t) = i \ld \m Z(t) [\m P,\m
O]. \label{fback} \ee If the linear control operator $\m P$ comes
from the mechanical oscillator, we have $i[\m P, q]=const.$ and $[\m
P, F]=0$. In the frequency domain, Eq. (\ref{fback}) gives \be
q_{fb}(\og)&=& i\ld' \m Z(\og)/\og, \nn F_{fb}(\og)&=&0, \quad \ld'=
i[\m P, q]\ld. \label{fback2} \ee Combing Eqs. (\ref{lre}) and
(\ref{fback2}), the final equation of motion is obtained, \be
q(\og)&=& q_0(\og)+ \chi_{qx}(\og)f(\og)\nn && + g \chi_{qq}(\og)
F(\og)+i \ld' \m Z(\og)/\og, \nn F(\og) &=& F_0(\og)+ g
\chi_{F\!\!\!\ F}(\og) q(\og), \nn \m Z(\og) &=& \m Z_0(\og)+ g
\chi_{\m Z F}q(\og). \ee It is checked that the force estimator
$\hat f$ deduced from the above equation is identical with Eq.
(\ref{add2}). Similar result can be obtained if the control operator
$\m P$ is from the detector. Therefore, the generalized UQL is valid
in the presence of the direct quantum feedback control.

\end{document}